\documentclass[aps,preprint]{revtex4}

\usepackage{graphicx}

\begin{document}

%\draft

\title{The bulk viscosity of a symmetrical Lennard--Jones mixture above and at
liquid--liquid coexistence: A computer simulation study}
\author{Subir K. Das, J\"urgen Horbach, and Kurt Binder}
\affiliation{Institut f\"ur Physik, Johannes Gutenberg--Universit\"at,
Staudinger Weg 7, D--55099 Mainz, Germany}

\begin{abstract}
A Lennard--Jones model of a binary dense liquid (A,B) with a symmetrical
miscibility gap is investigated by means of computer simulation methods.
Semigrand--canonical Monte Carlo simulations yield the phase diagram
in the $T$--$x$ plane ($T$: temperature, $x$: concentration of A or
B particles) as well as equilibrated configurations at coexistence.
Then Molecular Dynamics simulations use these configurations to
determine static properties (isothermal compressibility $\kappa_T$
and concentration susceptibility $\chi$) as well as the shear and the bulk
viscosity $\eta_{\rm s}$ and $\eta_{\rm B}$, respectively. The latter
quantities are calculated along a path approaching the coexistence line
from high temperatures in the one--phase region and ending at a state
at the coexistence line about 15\% below the critical point.  We find
that $\kappa_T$ and $\chi$ increase significantly near the coexistence
line reflecting the vicinity of the critical point. Whereas $\eta_{\rm
s}$ exhibits a weak temperature dependence, $\eta_{\rm B}$ increases
significantly near the coexistence curve.
\end{abstract}

\maketitle

\section{Introduction}
\label{sec1}

The bulk viscosity (in the following denoted by $\eta_{\rm B}$) describes
the response of a fluid to a compression or expansion. Compared to
other transport coefficients such as the shear viscosity or the self
diffusion constant, it is the least studied transport coefficient. This is
surprising since $\eta_{\rm B}$ is for instance a central quantity in the description
of the damping of longitudinal sound. It is also an important
quantity to probe slow dynamic processes such as the critical slowing down
near the critical point of a liquid--gas transition or the liquid--liquid
unmixing transition in a binary fluid. We will briefly discuss these
issues below.

A microscopic expression for $\eta_{\rm B}$ is given by a Green--Kubo 
formula~(Boon and Yip, 1980),
\begin{equation}
  \eta_{\rm B} = \frac{V}{k_B T} \int_0^{\infty} \;
         \left< J_{\alpha \alpha}(t) J_{\alpha \alpha}(0) \right> \ ,
  \label{eq_i_1}
\end{equation}
with $\alpha$ denoting Cartesian components ($\alpha \in \{x, y, z\}$).
$V$, $T$ and $k_B$ are volume, temperature and Boltzmann's constant,
respectively.  In the microcanonical ensemble, $J_{\alpha \alpha}$
is equal to the difference between the pressure at time $t$ and that
at $t=0$, $J_{\alpha \alpha} (t) = p(t) - p(0)$, where $p(t)$ is equal
to the diagonal elements of the pressure tensor ${\bf \sigma}$ defined
as follows:
\begin{equation}
   \sigma_{\alpha \beta} =
   \frac{1}{V} \, \sum_{i=1}^{N} \left[
    m_i v_{i \alpha} v_{i \beta} +
    r_{i \alpha} F_{i \beta} \right] \ . 
  \label{eq_i_2}
\end{equation}
Herein $m_i v_{i \alpha}$ and $r_{i \alpha}$ are
respectively the $\alpha$'th component of momentum and position of
particle $i$, and $F_{i \alpha}$ is the $\alpha$'th component of the
force acting on particle $i$.  Note that in order to calculate the shear
viscosity $\eta_{\rm s}$ one has to use the non--diagonal elements of the pressure
tensor in the Green--Kubo integral ($\alpha \neq \beta$)~(Boon and Yip, 1980):
\begin{equation}
   \eta_{\rm s} = \frac{V}{k_B T} \int_0^{\infty} dt \;
           \left< \sigma_{\alpha \beta} (t) 
           \sigma_{\alpha \beta} (0) \right>  \ .
  \label{eq_i_3}
\end{equation}
Eqs.~(\ref{eq_i_1}), (\ref{eq_i_2}), and (\ref{eq_i_3}) can be used to
calculate $\eta_{\rm B}$ and $\eta_{\rm s}$ from equilibrium fluctuations in a
Molecular Dynamics (MD) computer simulation. Indeed, in one of the
pioneering MD studies of a Lennard--Jones liquid near its triple point by
Levesque {\it et al.}~(Levesque et al., 1973) the viscosities were determined
by Green--Kubo formulas. Note that a recently proposed "new" formula
by Okumura and Yonezawa~(Okumura and Yonezawa, 2002) 
just expresses the pressure
fluctuations in Eq.~(\ref{eq_i_1}) in terms of the pair correlation
function and the interatomic potentials.

Alternative methods to determine $\eta_{\rm B}$ are based
on Non--Equilibrium Molecular Dynamics (NEMD) simulations.
Heyes~(Heyes, 1984; Heyes, 1986) proposed a NEMD scheme where the volume of
the system is changed from $V$ to $V+\Delta V$ at $t=0$ which leads to a
change of the pressure.  Then one follows the relaxation of the pressure
to its equilibrium value $p(\infty)$ and measures $p(t)-p(\infty)$
where $p(t)$ is the instantaneous pressure at time $t$. $\eta_{\rm B}$
is then given by
\begin{equation}
  \eta_{\rm B} = - \frac{V}{\Delta V} \int_0^{\infty} 
                 [ p(t) - p(\infty) ] dt  \ .
  \label{eq_i_4}
\end{equation}
Certainly, Eq.~(\ref{eq_i_4}) is only valid if $\Delta V$ is small enough
to allow the application of linear response theory.

Another NEMD approach was proposed by Hoover {\it et al.}~(Hoover et al., 1980).
The latter authors impose a frequency--dependent small perturbation
by changing the volume of the system by a periodic compression and
expansion with a frequency $\omega$. As a result $\eta_{\rm B}(\omega)$
is obtained for several values of $\omega$ and then an extrapolation to
zero frequency may be possible. The method of Hoover {\it et al.} might
be especially useful for liquid states where $\eta_{\rm B}$ exhibits
a long--time tail. However, some knowledge of the frequency dependence
of $\eta_{\rm B}(\omega)$ is required to extrapolate it accurately from
finite frequencies to zero.

The aforementioned methods have been mainly used in feasibility
studies where the bulk viscosity was determined, e.g., for a
Lennard--Jones fluid at a single state near its triple point
(see Refs.~(Levesque et al., 1973; 
Hoover et al., 1980; Heyes, 1984; Okumura and Yonezawa, 2002)).
Only in a small number of simulations, $\eta_{\rm B}$ has been
investigated systematically.  One of these rare studies is the MD
simulation of symmetrical Lennard--Jones mixtures by Vogelsang and
Hoheisel~(Vogelsang and Hoheisel, 1988; Hoheisel, 1993) who considered systems of
256 particles at moderate densities (i.e.~far from the triple point).
In this work $\eta_{\rm B}$ as well as $\eta_{\rm s}$ were calculated by means
of the Green--Kubo formulas, Eqs.~(\ref{eq_i_1}) and (\ref{eq_i_3}).
An interesting result of this study was that the ratio $\eta_{\rm
B}/\eta_{\rm s}$ is (much) larger than one if the fluid mixture has a (strongly)
associating character or a (strongly) demixing character. In both of
the latter cases the bulk viscosity increases quickly whereas the shear
viscosity remains essentially constant. As a consequence it is expected
that $\eta_{\rm B}$ shows a strong increase near the coexistence line
of a fluid--fluid unmixing transition.

In contrast to the small number of simulations, there are
many theoretical investigations of the bulk viscosity in the
context of the dynamics near the liquid--gas critical point
(Kawasaki, 1976; 
Kadanoff and Swift, 1968; Swift, 1968; Hohenberg and Halperin, 1977; 
Folk and Moser, 1995; Onuki, 1997; Onuki, 2002).
These works predict that the bulk viscosity exhibits a strong
divergence near the critical point of a gas--liquid liquid transition.
In contrast to that, the shear viscosity is expected to show a very weak
divergence (logarithmic divergence) at the critical point (if there is
at all a divergence in this quantity).  The latter predictions have been
confirmed experimentally.  An example is $^3$He in the vicinity of the
critical temperature $T_{\rm c}$: At $T/T_{\rm c} - 1 = 10^{-4}$ on the
critical isochore, $\eta_{\rm B}$ is about 50 Poise whereas $\eta_{\rm s}$
is equal to $17 \times 10^{-6}$ Poise (Kogan and Meyer, 1998; Onuki, 2002).

In the present work we consider a simple model of a {\it dense} liquid
mixture near and at a liquid--liquid unmixing transition and,
apart from static susceptibilities, we calculate the shear and the bulk viscosity.
Although we are not able to determine these quantities very close to
the critical point, we find a behavior which agrees qualitatively
with the aforementioned theoretical predictions for the critical
dynamics: $\eta_{\rm B}$ shows a stronger increase than $\eta_{\rm s}$
when approaching a state on the coexistence line about 15\% below the
critical point and, furthermore, at the latter point, $\eta_{\rm B}$ is
significantly larger than $\eta_{\rm s}$, i.e.~$\eta_{\rm B}/\eta_{\rm
s}\approx 3.3$.

The rest of the paper is organized as follows: In the next section
we briefly comment on the details of the simulation as well as the
Lennard--Jones model and its phase diagram. The static properties and the
transport coefficients (shear and bulk viscosity) as obtained from the
simulation are then presented in Sec.~3. Finally we summarize the 
results in Sec.~4.

\section{Model and Phase Diagram}
\label{sec2}

The model that we consider in this work is a binary Lennard--Jones
mixture. Thus, the interaction potential between a particle of type
$\alpha$ and a particle of type $\beta$ ($\alpha, \beta \in \{ {\rm A,
B} \}$) is given by
\begin{equation}
u_{\alpha \beta}(r)=
4 \epsilon_{\alpha \beta} 
\left[ \left( \frac{\sigma_{\alpha \beta}}{r} \right)^{12} -
       \left( \frac{\sigma_{\alpha \beta}}{r} \right)^6
\right] \ ,
\label{eq1}
\end{equation}
$r$ being the distance between the two particles.  For the Lennard--Jones
parameters $\epsilon_{\alpha \beta}$ and $\sigma_{\alpha \beta}$ we
choose $\sigma_{\rm AA} = \sigma_{\rm BB} = \sigma_{\rm AB} = \sigma$,
$\epsilon_{\rm AA} = \epsilon_{\rm BB} = \epsilon$ and $\epsilon_{\rm
AB}= \delta \epsilon$. Lengths, energies, and temperatures are measured
respectively in units of $\sigma \equiv 1$, $\epsilon \equiv 1$, and
$\epsilon/ k_B \equiv 1$. In the Molecular Dynamics (MD) part equal
masses are chosen for A and B particles, i.e.~$m_{\rm A} = m_{\rm B} =
1$. The potential is truncated and shifted at $r=2.5 \sigma$.

The model mixture that we have defined so far is obviously completely
symmetrical. Whether it has the tendency towards association or demixing
is controlled by the parameter $\delta$. We use $\delta = 0.5$ which
implies the possibility of a fluid--fluid unmixing transition.  Since we
are interested in the dense liquid state we have chosen a density $\rho
\sigma^3 = 1$, which provides the absence of crystallization in the
temperature range of interest, $T > 1.0$. Note that for densities $0
\le \rho \le 0.7$ the phase behavior of symmetrical LJ mixtures have
been extensively investigated by Wilding~(Wilding, 1997).

The simulations were done as follows: We started with a random mixture
with an equal number of A and B particles. By using standard Monte Carlo
(MC) in the canonical ensemble with trial displacements of particles
in the range $[ -\sigma/20, +\sigma/20]$, we equilibrated the system
for $10^5$ Monte Carlo steps (MCS) per particle. Then, we switched
on a MC simulation in the semigrand--canonical ensemble, i.e., at the
end of each displacement sweep an identity switch of $N/10$ randomly
chosen particles was attempted, $A \to B$ or $B \to A$ ($N$ being the
total number of particles). Note that in the Metropolis criterion of
the semigrand--canonical moves, the chemical potential energy difference
$\pm \Delta \mu = \mu_{\rm A} - \mu_{\rm B}$ ($\mu_{\alpha}$: chemical
potential of species $\alpha \in \{ {\rm A, B} \}$) has to be taken
into account in addition to the energy change in the Boltzmann factor.
In order to localize the coexistence curve of the liquid--liquid unmixing
transition in the present case, one has to just set $\Delta \mu = 0$
which is simply due to the symmetry of our model.  In order to determine
the phase diagram we have performed five independent runs with a length
of 400000 MCS per particle where we started the averaging after 100000 MCS
in each run (for more details of this calculation see Ref.~(Das et al., 2003)).

\begin{figure}[ht]
\includegraphics[width=140mm]{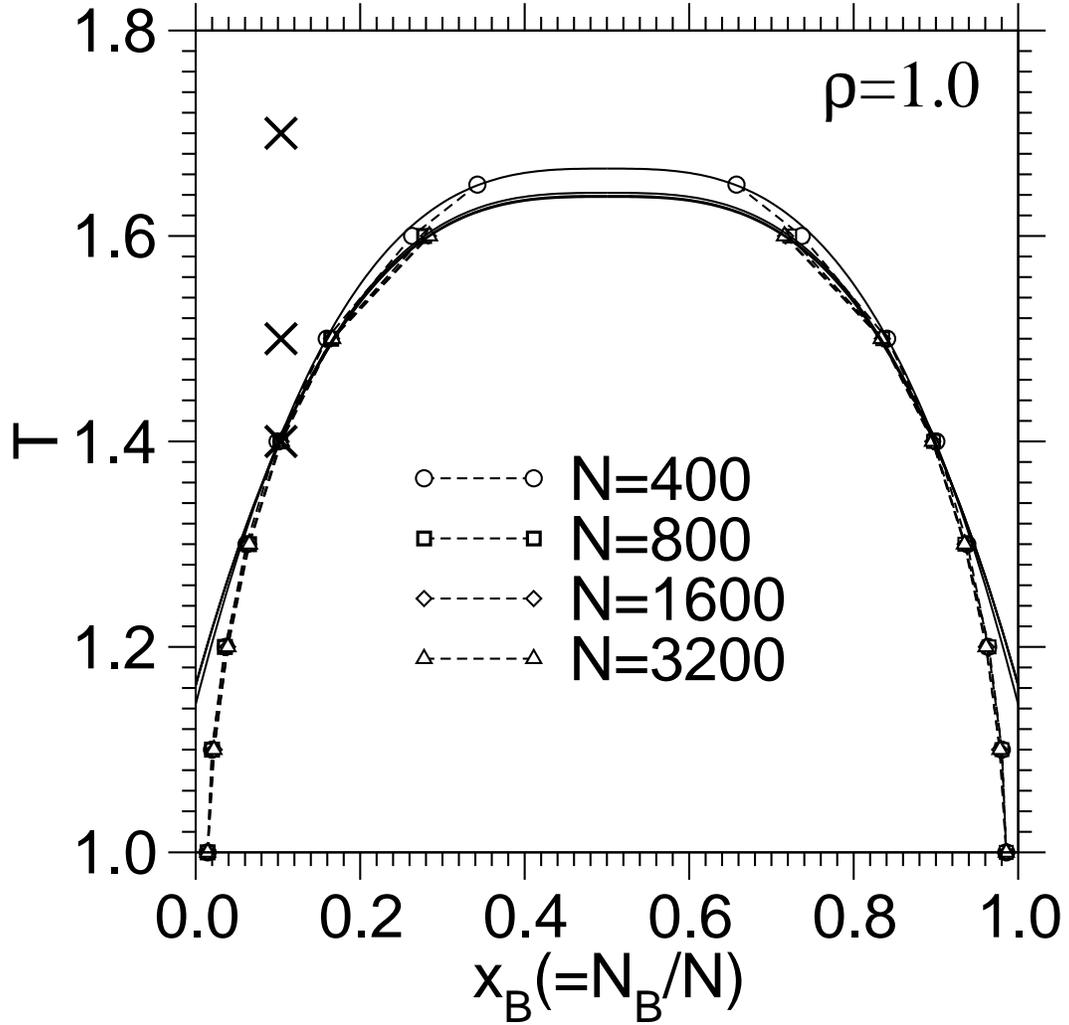}
\caption{Phase diagram of the symmetrical Lennard--Jones mixture for four
choices of $N$ as indicated. The crosses at $x_{\rm B}=0.10375$ mark the 
states for which the structure and dynamics was investigated (note that also
$T=3.0$ and $T=6.0$ were studied). 
The solid lines are fits with Eq.~(\ref{eq2}) and the dashed lines are 
just guides to the eye.
} 
\label{fig1}
\end{figure}

Fig.~\ref{fig1} shows the phase diagram in the $T$--$x_{\rm B}$ plane for
the system sizes $N=400, 800, 1600$, and 3200 ($x_{\rm B} \equiv N_{\rm
A}/N$ is the concentration of $B$ particles).  Due to the symmetry of
the model we know {\it a priori} that the critical point is located at
$x_{\rm B}=0.5$.  As we can infer from Fig.~\ref{fig1}, the finite size
effects near the critical point are small for $N \ge 400$, and for $N \ge
1600$ the data agree within the statistical errors. We have estimated
the critical temperature $T_{\rm c}$ from power law fits according to
the three--dimensional Ising universality class~
(Binder and Ciccotti, 1996; Binder and Heermann, 2002),
\begin{equation}
   f(x_{\rm B}) = 0.5 \pm x_{\rm B}^{\rm coex} = \hat{B}
                  \left( 1 - T/T_{\rm c} \right)^{\beta},
   \quad \quad  \beta \approx 0.325
 \label{eq2}
\end{equation}
where $\hat{B}$ is a critical amplitude which is used, as well as
$T_{\rm c}$, as a fitting parameter. From the fits with Eq.~(\ref{eq2})
we obtain $T_{\rm c} \approx 1.638 \pm 0.005$ for $N\ge 1600$. For a
more accurate estimate of $T_{\rm c}$, we would have to perform a 
finite scaling analysis~(Binder and Ciccotti, 1996; 
Binder and Heermann, 2002).

Apart from the phase diagram the MC in the semigrand--canonical ensemble
yields also equilibrated configurations exactly along the coexistence
line. We used them as starting configurations for Molecular Dynamics
(MD) simulations to determine the static quantities and the transport
coefficients that are presented in the next section. In the MD, the
equations of motion were integrated by means of the velocity Verlet
algorithm~(Binder and Ciccotti, 1996) with a time step $\delta t = 0.01$ [in units
of the time $t_0 = (m \sigma^2/(48 \epsilon))^{1/2}$].

Starting point for the MD were the configurations with 1600 particles
at $T=1.4$ that correspond to the concentration $x_{\rm B}=0.10375$ at
the coexistence line.  Configurations in the one--phase region at the
latter value of $x_{\rm B}$ were obtained by heating up the system and
equilibrating it for $10^5$ time steps at constant temperature with the
use of an Andersen thermostat~(Das et al., 2003). Then, microcanonical runs were added from
which we computed the static and dynamic quantities that are shown in the
next section. The path along which we determined the latter quantities is
indicated in Fig.~\ref{fig1} by crosses: Apart from the coexistence state
at $T=1.4$, which is about 15\% below the critical point with respect
to temperature, the temperatures $T=1.6$, 1.7, 3.0, and 6.0 were analyzed
(note that also other paths around the coexistence line are studied in
Ref.~(Das et al., 2003)).

One may wonder why we have not studied states that are much closer
to the critical point. But due to the diverging correlation length
that is accompanied by the approach of the critical point, we would
have to consider systems that contain much more than 1600 particles
as in our work.  Furthermore, the critical slowing down would require
very long runs to equilibrate the system and to determine the transport
coefficients with reasonable error bars. The latter point is especially a
severe problem for transport coefficients such as the shear or the bulk viscosity.
These are collective quantities and require many independent runs and/or
a long time averaging since they are not subject to a self--averaging as
one--particle quantities such as the self--diffusion constant. However,
compared to many previous works, our choice of $N$ is relatively
large. Even the very recent computation of the bulk viscosity
by Okumura and Yonezawa~(Okumura and Yonezawa, 2002) 
was only done for a small
system of $256$ particles.

\section{Results}
\label{sec3}

In this section we present the results for the static and dynamic
properties of the symmetrical LJ system along the path in the phase
diagram which is indicated in Fig.~\ref{fig1}.  As described in the
previous section, we have generated first five independent configurations
at each temperature.  All these configurations were used as initial
configurations for microcanonical MD runs over 4.8 million time steps
(the time step was 0.01 $t_0$, see previous section). Thus, at each
temperature 24 million time steps were done to determine the quantities
of interest. As we shall see in the following, this effort was large
enough to get a reasonable estimate of bulk and shear viscosities.

\subsection{Static Properties}

As we see in Fig.~\ref{fig1} the states at $T=1.4$ on the coexistence line
are about 15\% below the critical point with respect to temperature.
Although these points are not very close to the critical point one may expect
that the approach of the critical point is reflected in thermodynamic
quantities such as the isothermal compressibility $\kappa_T$ and the
concentration susceptibility $\chi$ (defined below).

$\kappa_T$ can be calculated from the static number--number density structure
factor $S_{\rm nn}(q)$ in the limit of 
wavenumber $q\to 0$~(Hansen and McDonald, 1986),
\begin{equation}
   \kappa_T = \frac{1}{\rho k_B T} \lim_{q\to 0} S_{\rm nn}(q)
  \label{eqsec3_0}
\end{equation}
with $\rho$ being the total density of the system (in our case $\rho$
as well as the Boltzmann constant $k_B$ are equal to one). The structure factor 
$S_{\rm nn}(q)$ for a binary AB mixture is defined by~(Hansen and McDonald, 1986)
\begin{equation}
  S_{\rm nn}(q) = S_{\rm AA}(q) + 2 S_{\rm AB}(q) + S_{\rm BB}(q)
  \label{eqsec3_1}
\end{equation}
where $S_{\alpha \beta}(q)$ ($\alpha, \beta \in [A,B]$) are the partial 
structure factors,
\begin{equation}
   S_{\alpha \beta}(q) = \frac{f_{\alpha \beta}}{N}
                \sum_{i,j} \left< 
                 \exp \left[ i {\bf q} \cdot ({\bf r}_i^{\alpha} 
                        - {\bf r}_j^{\beta})
                   \right] \right>
  \label{eqsec3_2}
\end{equation}
with $f_{\alpha \beta}=0.5$ for $\alpha \neq \beta$ and $f_{\alpha
\beta}=1.0$ for $\alpha = \beta$. In Eq.~(\ref{eqsec3_2}) the indices
$i,j$ run over the number of particles of species $\alpha$ and $\beta$,
respectively, and ${\bf r}_i^{\alpha}$ is the position of the $i$'th
particle of species $\alpha$.

\begin{figure}[ht]
\includegraphics[width=140mm]{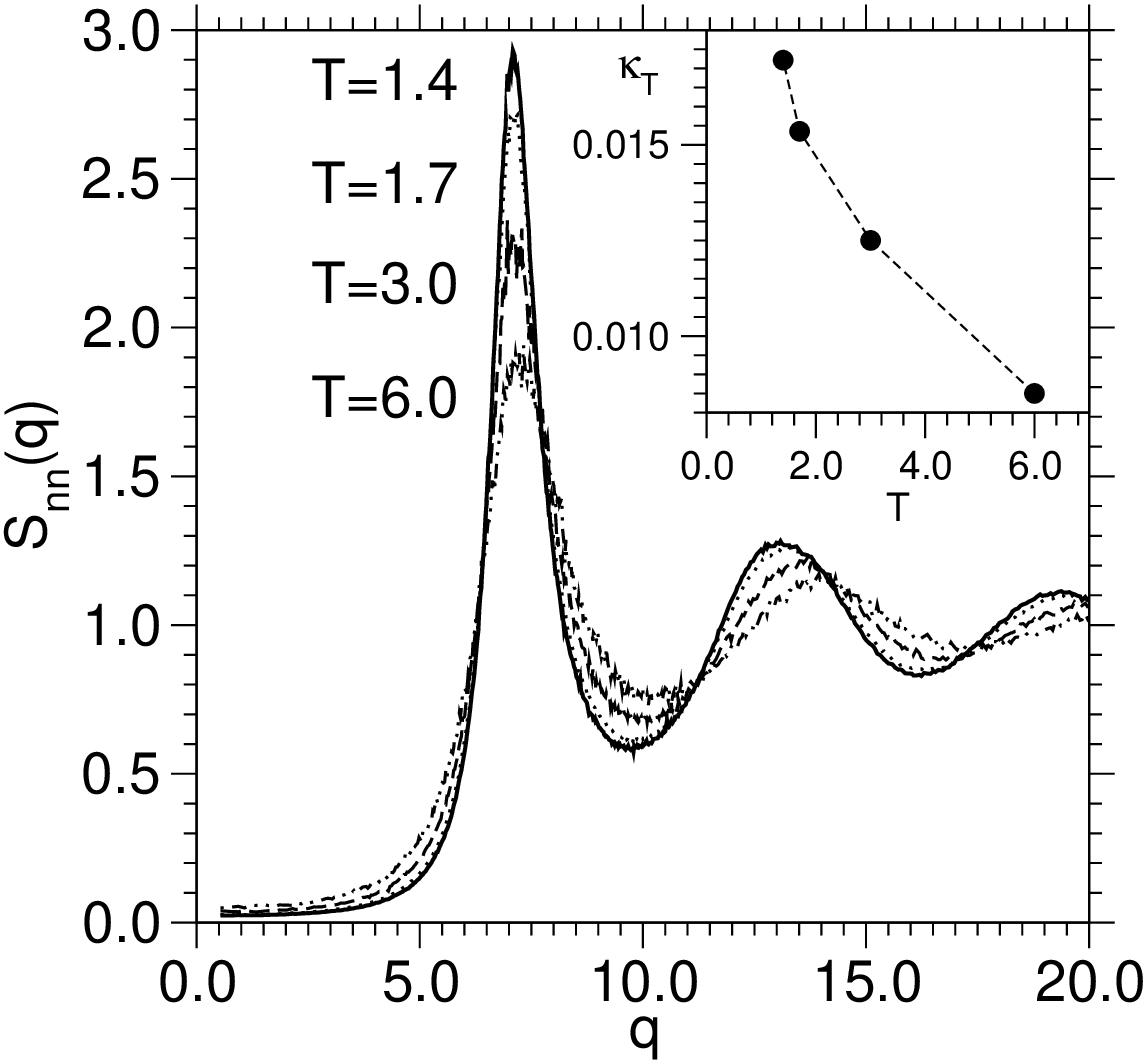}
\caption{
Number--number density structure factor $S_{\rm nn}(q)$ for the four 
indicated temperatures. The inset shows the isothermal compressibility
$\kappa_T$ as a function of temperature. $\kappa_T$ is estimated from
the extrapolated value $S_{\rm nn}(q=0)$ [see Eq.~(\ref{eqsec3_0})].
} 
\label{fig2}
\end{figure}

Fig.~\ref{fig2} shows $S_{\rm nn}(q)$ for $T=1.4$, 1.7, 3.0, and 6.0.
For wavenumbers $q$ that correspond to distances smaller or equal the
typical nearest neighbor distance, say $q>5$, the typical behavior of
this quantity for simple dense liquid can be identified: Upon decreasing
the temperature the amplitude especially of the first peak increases
and the peaks become narrower.  The small values of $S_{\rm nn}(q)$
for $q\to 0$ reflect the fact that the considered dense liquid state is hardly
compressible. It might be surprising that even at coexistence $S_{\rm
nn}$ does not show any tendency to increase significantly for $q\to 0$.
The amplitude of $S_{\rm nn}(q)$ at small $q$ appears to be 
even a monotonic decreasing
function with decreasing temperature. However, the relevant thermodynamic
quantity in our context is $\kappa_T$, that we have extracted from
$S_{\rm nn}(q)$ by extrapolating this function to $q=0$. As we see in
the inset of Fig.~\ref{fig2}, $\kappa_T$ increases significantly with
decreasing temperature which shows that for states around $T=1.4$,
long--ranged static correlations, i.e.~the presence of the critical
point, still affect the behavior of $\kappa_T$.

\begin{figure}[ht]
\includegraphics[width=140mm]{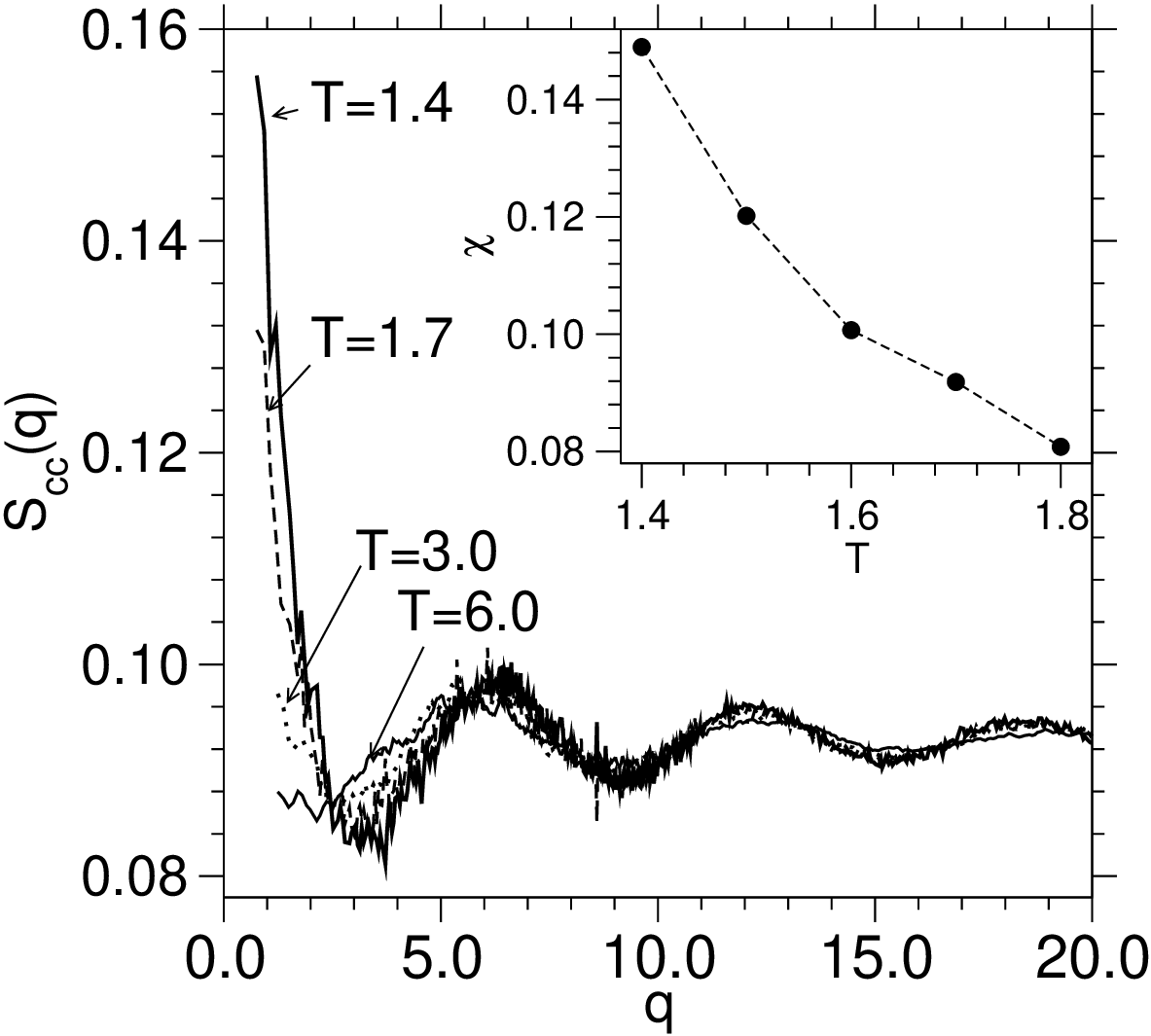}
\caption{
Concentration--concentration density structure factor $S_{\rm cc}(q)$ for the
four indicated temperatures. The inset shows the concentration susceptibility
$\chi$ as function of temperature (see text).
} 
\label{fig3}
\end{figure}

The "vicinity" of the critical point is more pronounced in the structure
factor of the concentration densities, $S_{\rm cc}(q)$, than in $S_{\rm
nn}(q)$.  $S_{\rm cc}(q)$ can be also expressed by a linear combination
of the partial structure factors~(Hansen and McDonald, 1986), i.e.
\begin{equation}
  S_{\rm cc}(q) = x_{\rm B}^2 S_{\rm AA}(q) 
        - 2 x_{\rm A} x_{\rm B} S_{\rm AB}(q) 
        + x_{\rm A}^2 S_{\rm BB}(q) \ .
  \label{eqsec3_3}
\end{equation}
In the limit $q \to 0$ the structure factor $S_{\rm cc}(q)$ is related to
the static concentration susceptibility $\chi$ by
\begin{equation}
   \chi = \frac{1}{k_B T} \lim_{q\to 0} S_{\rm cc}(q) \ .
\end{equation}
Note that we have determined $\chi$ directly via a fluctuation relation
by semigrand--canonical MC runs (see Ref.~(Das et al., 2003)). So, it was
not necessary to extrapolate $S_{\rm cc}(q)$ to $q=0$. As we see in
Fig.~\ref{fig3} this would be a difficult task because, in contrast to
$S_{\rm nn}(q)$, $S_{\rm cc}(q)$ steeply increases for $q \to 0$. 
As we can infer from the inset of Fig.~\ref{fig3}, $\chi$ is about a
factor of 2 larger at the coexistence state at $T=1.4$ than at $T=1.8$.
It is remarkable that $S_{\rm cc}(q)$ exhibits almost no temperature
dependence for $q>5$ in the broad temperature range $1.4 \le T \le 6.0$.

\subsection{Bulk viscosity and shear viscosity}

For the computation of the bulk and shear viscosities we have used
the Green--Kubo (GK) formulas, Eqs.~(\ref{eq_i_1}) and (\ref{eq_i_3}).
The alternative methods that are based on NEMD require essentially
the same computational effort. Furthermore, in the Heyes method, see
Eq.~(\ref{eq_i_4}), one has to choose the perturbation $\Delta V$ small enough
to ensure that this perturbation justifies the application of linear
response theory. Thus, one has to study the dependence of the measured
bulk viscosity $\eta_{\rm B}$ on $\Delta V$ (of course, in the linear
response regime the apparent $\eta_{\rm B}$ is independent of 
$\Delta V$).  The Hoover method has in
addition the drawback that one has to extrapolate the frequency--dependent
viscosity to zero frequency. However, a comparative study of the different
NEMD and GK methods to measure transport coefficients in a simulation
is an interesting future project since the NEMD methods may give additional
physical insight into the microscopic mechanism of different transport
processes.

\begin{figure}[ht]
\includegraphics[width=140mm]{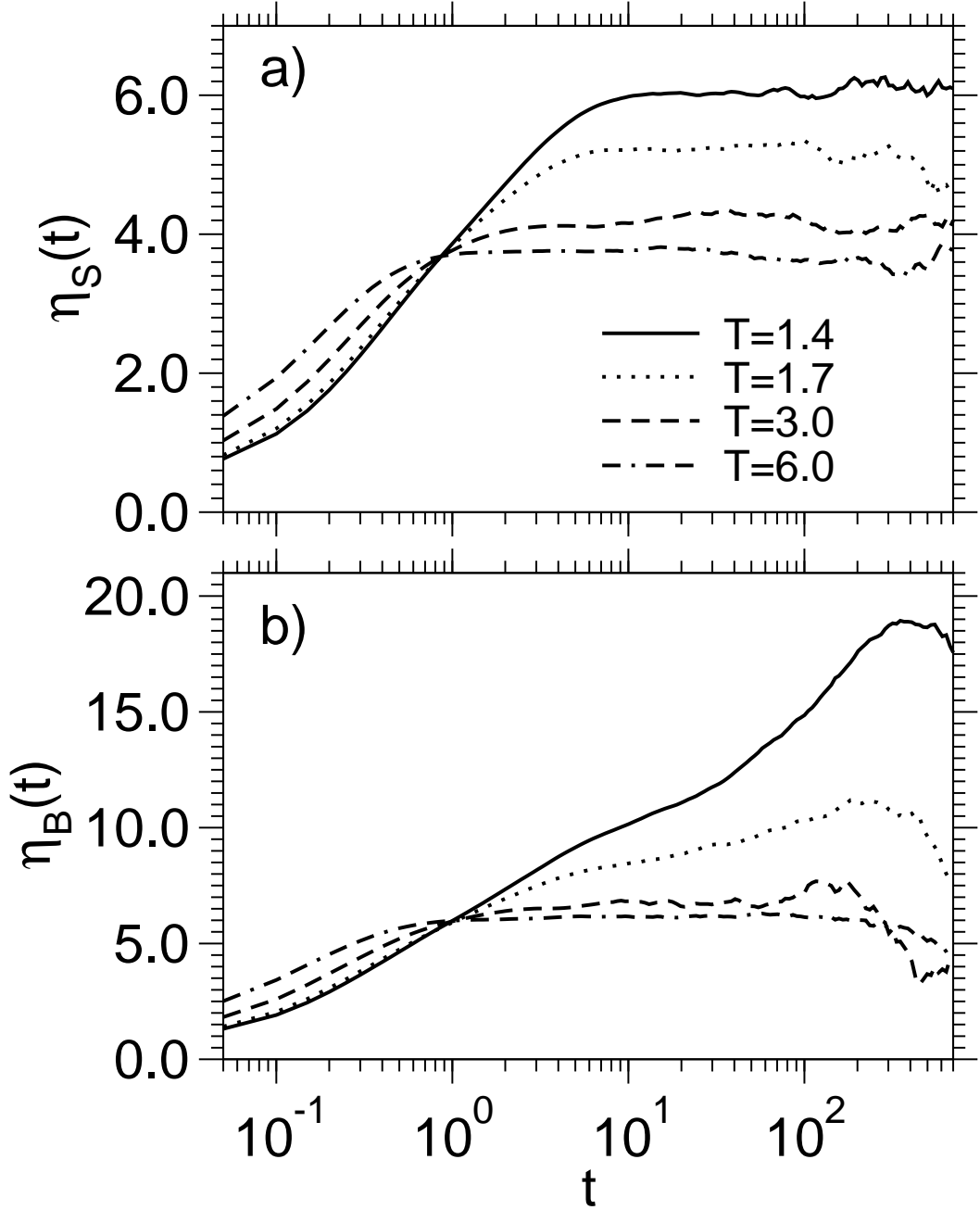}
\caption{
a) "Time--dependent" shear viscosity $\eta_{\rm s}(t)$ for the 
   indicated temperatures. From the long--time plateau we read off
   $\eta_{\rm s}$. 
b) Same as in a), but now for the bulk viscosity.
} 
\label{fig4}
\end{figure}

Fig.~\ref{fig4} shows $\eta_{\rm s}(t)$ and $\eta_{\rm B}(t)$ for four
temperatures. These quantities are defined by Eqs.~(\ref{eq_i_1}) and
(\ref{eq_i_3}) where one has to replace $\infty$ in the integral by $t$. We
see that $\eta_{\rm s}(t)$ and $\eta_{\rm B}(t)$ approach indeed plateaus
at long times the values of which correspond to the hydrodynamic shear
and bulk viscosities, respectively.  At low temperatures, there is a
qualitative difference in $\eta_{\rm B}(t)$ as compared to $\eta_{\rm
s}(t)$: E.g.~at $T=1.4$, $\eta_{\rm s}(t)$ is essentially constant
for $t>10$. In contrast to that, $\eta_{\rm B}(t)$ exhibits a second
strong increase and it reaches the plateau value for $t>300$. This is
due to a long--time tail in the autocorrelation function of the pressure
fluctuations.  Note that the decrease of $\eta_{\rm B}(t)$ for $t>500$
is due to the fact that the statistics is much worse at long times.

\begin{figure}[ht]
\includegraphics[width=120mm]{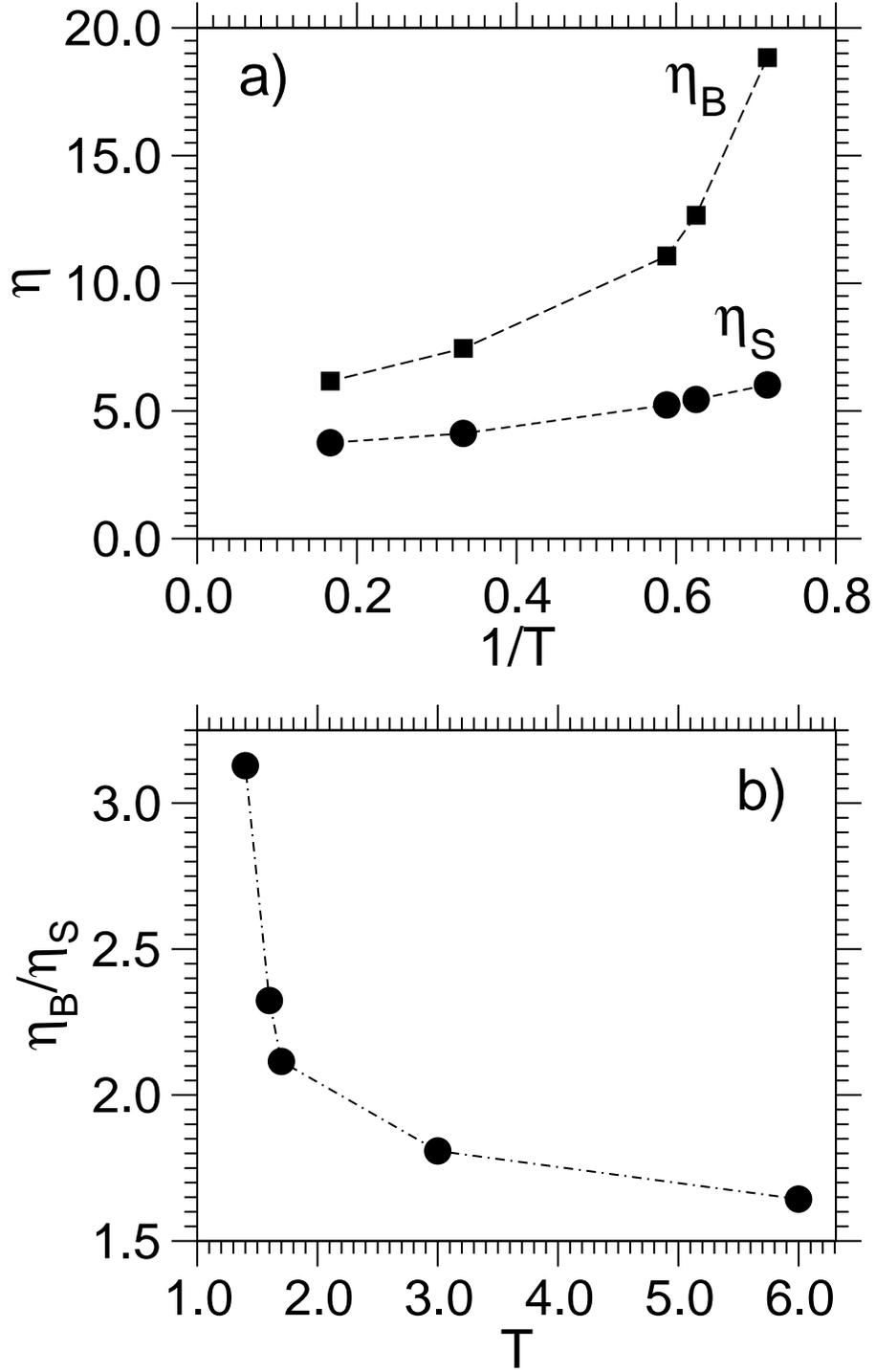}
\caption{
a) Shear and bulk viscosity as a function of inverse temperature.
b) Ratio $\eta_{\rm B}/\eta_{\rm s}$ as a function of temperature.
} 
\label{fig5}
\end{figure}

$\eta_{\rm B}$ and $\eta_{\rm s}$ are plotted in Fig.~\ref{fig5}a as
a function of inverse temperature. Whereas $\eta_{\rm s}$ exhibits
only a very weak temperature dependence, $\eta_{\rm B}$ increases
significantly in the vicinity of the coexistence state at $T=1.4$.
As we can see in Fig.~\ref{fig5}b the ratio $\eta_{\rm B}/\eta_{\rm s}$
is in the whole considered temperature range $6.0 \ge T \ge 1.4$ above
one, and it reaches a value of about 3.3 at $T=1.4$. One expects such
a behavior from theories of the critical dynamics of the liquid--gas
transition~(Onuki, 2002). According to these theories the long--ranged
critical fluctuations cause a slowing down of the system's response to
a compression or expansion (described by $\eta_{\rm B}$). On the other
hand, the response to the shearing of the system is hardly affected by
the critical fluctuations (and thus $\eta_{\rm s}$). In our case, at a
state about 15\% below the critical point, there is already a significant
increase of static correlations which makes the behavior of $\eta_{\rm
B}/\eta_{\rm s}$ plausible.

Since the data presented in this paper are taken at an off-critical
concentration, one could also attempt to interpret them in terms of
a singular behavior at the ``spinodal temperature'' $T_{\rm s}$ (limit of
metastability) (Binder, 1987). According to the mean field theory of
symmetric binary mixtures, one should expect that the static
concentration susceptibility $\chi$ for $x_{\rm B} < x_{\rm B}^{\rm crit} 
= 0.5$ behaves as
\begin{equation}
\chi(T,x_{\rm B}) \propto [T-T_{\rm s}(x_{\rm B})]^{-1}
\label{eqn12}
\end{equation}
where near the critical temperature the spinodal temperature $T_{\rm s}
(x_{\rm B})$
is the inverse function of the concentration $x_{\rm B}^{\rm s}(T)$ along the
spinodal curve, given by the equation
$x_{\rm B}^{\rm s}(T)-x_{\rm B}^{\rm crit}=(x_{\rm B}^{\rm coex}-
x_{\rm B}^{\rm crit})/\sqrt 3$ (Binder, 1987).
Further away from $T_{\rm c}$, a simple expression for $T_{\rm s}
(x_{\rm B})$ exists
for the lattice (Ising) model of symmetric binary mixtures, namely
\begin{equation}
x_{\rm B}^{\rm s}(T)=[1\pm \sqrt{1-T/T_{\rm c}^{\rm MF}}]/2.
\label{eqn13}
\end{equation}
Here we have emphasized by this notation that the mean field estimate
$T_{\rm c}^{\rm MF}$ of the critical temperature for systems with short range
forces normally exceeds the actual critical temperature distinctly
(also Eq.~(\ref{eqn12}) does then not hold for $x_{\rm B}$ near $x_{\rm B}^{
\rm crit}$
and $T$ near the actual critical temperature, since
$\chi(T,x_{\rm B}^{\rm crit})\propto (T-T_{\rm c})^{-\gamma}$, where the actual
susceptibility exponent $\gamma \approx 1.24$ (Binder and Ciccotti,
1996; Binder and Heermann, 2002)).

\begin{figure}[ht]
\includegraphics[width=140mm]{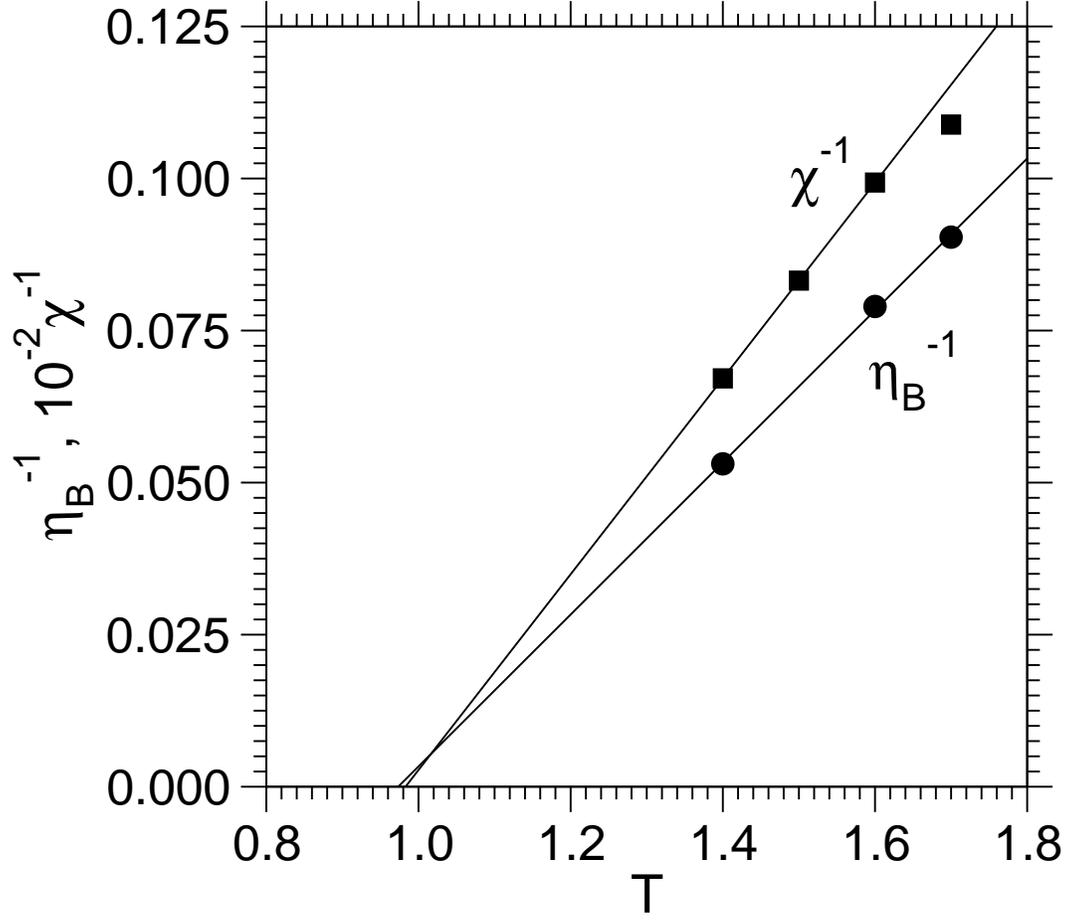}
\caption{
Mean-field type extrapolation towards the ``spinodal''. The solid lines
are fit to the data sets by using the functional form given by 
Eq.~(\ref{eqn12}).
}
\label{fig6}
\end{figure}

Although we do not really expect that Eq.~(\ref{eqn12}) and a
related mean-field divergence for the bulk viscosity $\eta_{\rm B}$ is
a good approximation for our Lennard-Jones system, we present
a plot of $\chi^{-1}$ vs. $T$ and $\eta_{\rm B}^{-1}$ vs. $T$ in Fig~\ref{fig6}.
Mean field theory would predict that the data fall on straight lines
and both straight lines should hit the abscissa in the same 
point which then is the estimate of $T_{\rm s}(x_{\rm B})$. Indeed the data
points close to the coexistence curve are compatible with such
analysis, with $T_{\rm s}(x_{\rm B})\approx 1$. Of course, one should not 
put too much weight on this analysis, since the temperature range
over which we need to extrapolate 
is larger than the temperature range where actual data are fitted.
Also the estimate from Eq.~(\ref{eqn13})
would be much lesser, namely $T_{\rm s}(x_{\rm B}) 
\simeq 0.59$, if the distinction
between the actual $T_{\rm c}$ and $T_{\rm c}^{\rm MF}$ is ignored. We
caution the reader that anyway the concept of a spinodal is of
doubtful validity outside of mean field theory (Binder,1987),
although in the experimental literature on binary mixtures
(both in metallic alloys and in polymer blends, for instance)
it is widely used.

\section{Summary}
\label{sec4}

We have used computer simulations to investigate transport coefficients
of a dense symmetrical Lennard--Jones mixture that were calculated along
a path towards a liquid--liquid miscibility gap ending at a coexistence
state about 15\% below the critical point. The main result of our study
is that the bulk viscosity $\eta_{\rm B}$ increases significantly near the
coexistence state whereas the shear viscosity $\eta_{\rm s}$ does not show
any change near coexistence.  $\eta_{\rm s}$ remains to exhibit a very
weak temperature dependence also when it passes the coexistence line. The
behavior of $\eta_{\rm B}$ and $\eta_{\rm s}$ can be qualitatively
understood by theories of critical dynamics (see Ref.~(Onuki, 2002)). 

In future studies we plan to compute the transport properties also closer
to the critical point.  Of course, in such studies much larger system
sizes than those used in this work have to be considered. Moreover,
the emergence of critical slowing down requires simulations on much 
longer time scales.

\noindent\\
Acknowledgement

\noindent\\ 
The present research was supported by the Deutsche Forschungsgemeinschaft
(DFG) under Grant No. Bi314/18 (SPP 1120). One of the authors
(J. H.) acknowledges the support of the DFG under Grants No. HO 2231/2-1
and HO 2231/2-2.

%\begin{thebibliography}{99}
\noindent\\
References

\begin{description}
\item[]
Binder, K. (1987). Theory of First Order Phase Transitions.
Rep. Prog. Phys., {\footnotesize 50}, 783.
\item[]
Binder, K. and Ciccotti, G., eds. (1996). Monte Carlo and Molecular Dynamics
of Condensed Matter Systems. Italian Physical Society, Bologna.
\item[]
Binder, K. and Heermann, D.W. (2002). Monte Carlo Simulations in Statistical
Physics. An Introduction, 4'th edition. Springer, Berlin.
\item[]
Boon, J.P. and Yip, S. (1980). Molecular Hydrodynamics. McGraw Hill, 
New York.
\item[]
Das, S.K., Horbach, J. and Binder, K. (2003). Transport phenomena and
microscopic structure in partially miscible binary fluids: A simulation
study of the symmetrical Lennard--Jones mixture. J. Chem. Phys. 
{\footnotesize 119}, 
1547.
\item[]
Folk, R. and Moser, G. (1995).
Nonuniversal Dynamical Crossover in Pure and Binary Fluids Near a Critical Point.
Phys. Rev. Lett. {\footnotesize 75}, 2706.
\item[]
Hansen, J.P. and McDonald, I.R. (1986). Theory of Simple Liquids.
Academic, London.
\item[]
Heyes, D.M. (1984). J. Chem. Soc., Faraday Trans. {\footnotesize 80}, 1363.
\item[]
Heyes, D.M. (1986). Can. J. Phys. {\footnotesize 64}, 774.
\item[]
Hoheisel, C. (1993). Theoretical Treatment of Liquids and Liquid Mixtures.
Elsevier, Amsterdam, pp. 292--300.
\item[]
Hohenberg, P.C. and Halperin, B.I. (1977). Theory of dynamic
critical phenomena. Rev. Mod, Phys. {\footnotesize 49}, 435.
\item[]
Hoover, W.G., Evans, D.J., Hickman, R.B., Ladd, A.J.C.,
Ashurst, W.T., and Moran, B. (1980). Lennard--Jones triple--point
bulk and shear viscosities. Green--Kubo theory, Hamiltonian mechanics,
and nonequilibrium molecular dynamics. Phys. Rev. A {\footnotesize 22}, 1690.
\item[]
Kadanoff, L.P. and Swift, J. (1968). Transport Coefficients near the
Liquid--Gas Critical Point. Phys. Rev. {\footnotesize 166}, 89.
\item[]
Kawasaki, K. (1976). Mode Coupling and Critical Dynamics. In:
Domb, C. and Green, M.S. (Eds.), {\it Phase Transitions and
Critical Phenomena}, Vol. 5A, pp. 166--405. Academic Press, London.
\item[]
Kogan, A.B. and Meyer, H. (1998). Sound Propagation in $^3$He and $^4$He Above
the Liquid--Vapor Critical Point. J. Low Temp. Phys. {\footnotesize 110}, 899.
\item[]
Levesque, D., Verlet, L., and K\"urkijarvi, J. (1973). Computer
"Experiments" on Classical Fluids. IV. Transport Properties
and Time--Correlation Functions of the Lennard--Jones Liquid
near Its Triple Point. Phys. Rev. A {\footnotesize 7}, 1690.
\item[]
Okumura, H. and Yonezawa, F. (2002). New formula for the bulk viscosity 
constructed from the interatomic potential and the pair distribution
function. J. Chem. Phys. {\footnotesize 116}, 7400.
\item[]
Onuki, A. (1997). Dynamic equations and bulk viscosity near the gas--liquid
critical point. Phys. Rev. E {\footnotesize 55}, 403.
\item[]
Onuki, A. (2002). Phase Transition Dynamics. Cambridge University Press, 
Cambridge,
p. 247.
\item[]
Swift, J. (1968). Transport Coefficients near the Consolute Temperature
of a Binary Liquid Mixture. Phys. Rev. {\footnotesize 173}, 257.
\item[]
Vogelsang, R. and Hoheisel, C. (1988). Thermal transport coefficients including
the Soret coefficient for various liquid Lennard--Jones mixtures.
Phys. Rev. A {\footnotesize 38}, 6296.
\item[]
Wilding, N.B. (1997). 
Critical end point behavior in a binary fluid mixture.
Phys. Rev. E {\footnotesize 55}, 6624.
%\bibitem{zaheri03}
%Zaheri, A. H. M., Srivastava, S., and Tankeshwar, K. (2003). Longitudinal
%and bulk viscosities of expanded rubidium. J. Phys.: Condens. Matter {\footnotesize 15},
%6683.
\end{description}
%\end{thebibliography}

\newpage

\end{document}